\documentclass{iopart}
\usepackage{iopams}  
\usepackage{color}
\usepackage{times}
\usepackage{graphicx}
\usepackage{fancyhdr}
\usepackage{float}

\def\thercsid{\relax}
\def\rcsid#1{\def\next##1#1{\def\thercsid{##1}}\next}
\rcsid$Id: cqg.tex,v 1.1 2004/04/20 19:06:46 patrick Exp $

\renewcommand{\today}{\number\day\space\ifcase\month\or
  January\or February\or March\or April\or May\or June\or
  July\or August\or September\or October\or November\or December\fi
  \space\number\year}

\begin{document}

\title{Incorporating information from source simulations into searches for
    gravitational-wave bursts}
\author{Patrick R Brady and Saikat Ray-Majumder}
\address{
    Dept. of Physics, PO Box 413\\
    University of Wisconsin-Milwaukee, WI 53201
}
\begin{abstract}
The detection of gravitational waves from astrophysical sources of
gravitational waves is a realistic goal for the current generation of
interferometric gravitational-wave detectors.    Short duration bursts
of gravitational waves from core-collapse supernovae or mergers of
binary black holes may bring a wealth of astronomical and
astrophysical information.   The weakness of the waves and the rarity
of the events urges the development of optimal methods to detect the
waves.   The waves from these sources are not generally
known well enough to use matched filtering however;  this drives
the need to develop new ways to exploit source simulation information
in both detections and information extraction.   We present an
algorithmic approach to using catalogs of gravitational-wave signals
developed through numerical simulation, or otherwise, to enhance our ability to
detect these waves.    As more detailed simulations become available,
it is straightforward to incorporate the new information into the search 
method.   This approach may also
be useful when trying to extract information from a gravitational-wave
observation by allowing direct comparison between the observation and
simulations.
\end{abstract}


\submitto{\CQG}

\section{Introduction}

The ability to extract weak signals from noise is generally enhanced
by having advanced knowledge of the expected signals~\cite{wainstein:1962}.  
For low-mass compact binary inspiral,
the expected waveform can be accurately computed in the band of
earth-based detectors,  and matched filtering
is used to search the detector output for the
signals~\cite{Allen:1999yt,Tagoshi:2000bz,Abbott:2003pj}.  These
searches for gravitational-wave chirp signals are the archetypical
example of using advanced knowledge of a waveform~\cite{Blanchet:1996pi} in
gravitational-wave astronomy.  Once the
signals are detected,   the extraction of information about the
astrophysical sources responsible for their generation will be of
great interest.   For waves from the inspiral of low-mass compact
binary systems,  the problem of information extraction is reasonably
straightforward.  A well developed formalism, based on matched
filtering, exists to characterize the accuracy of parameter estimation
for sufficiently strong signals~\cite{Cutler:ys,Finn:1992xs}.

In general,  the strongest sources of gravitational waves are composed
of high density material moving at relativistic speeds in extreme
gravitational fields and detailed computations of the expected
waveforms are not always possible.   Among the most plausible examples of
such sources are supernova
explosions~\cite{Cutler:2002me} or binary black hole
mergers~\cite{Flanagan:1997sx}.  Efforts to model these systems, and to
determine the gravitational waveforms emitted by them, are under
way~\cite{Zwerger:1997,Dimmelmeier:2001rw,Ott:2003qg,Lehner:2001wq,Baker:2002qf,Bruegmann:2003aw}.
As the simulations become more sophisticated,  the gravitational
waveforms become better approximations to the waves that will arrive
from real astrophysical systems.   A challenge for gravitational-wave
astronomy is to use the partial information from simulations to
enhance signal detection and information extraction.   If
robust features of the waves can be identified using the simulations 
and those features can be associated with
definite physical processes,  a great deal of information might be
obtained about the structure of dense matter or of spacetime itself.

Searches for gravitational-wave bursts have been made by a number of
groups~\cite{Nicholson:1996ys,IGEC:2000,Abbott:2003pb,igec:2003}.
These searches generally take an unbiased approach
by using time-frequency methods to identify excess power in the data
stream.  Plausible physical waveforms are only used to determine the efficiency
of the search and report an upper limit.  

In this note,  we outline a method to use information from numerical
simulations in searches for gravitational-wave bursts.   This method
is based on the excess-power statistic~\cite{Flanagan:1997sx,Anderson:2000yy} 
and provides an algorithmic approach to incorporating the
source-modeling information into searches.  Moreover,  it allows for
the incorporation of new information as it becomes available.  The
theoretical formulation of the method is presented in
Sec.~\ref{s:theory} with an example of the construction in
Sec.~\ref{s:example}.   Finally,  other possible applications are
mentioned in Sec.~\ref{s:summary}.

\section{Theory}
\label{s:theory}

The signal $s(t)$ from a gravitational wave detector is sampled to 
produce a time series $s_j=s(j\triangle t)$, where $j=0,1,2,\ldots,H-1$ and
$\triangle t$ is the time between samples. This signal can be written as
\begin{equation}
s_j=n_j+h_j
\label{e:signal-vector}
\end{equation}
where $n_j$ is the detector noise and $h_j$ is the (possibly absent)
gravitational-wave signal.  For the purposes of our presentation,  
we assume that the detector noise is
Gaussian with joint probability distribution
\begin{equation}
\frac{1}{\sqrt{(2\pi)^H \det||R||}} \exp \biggl\{ -\frac{1}{2} 
\sum_{j,k} n_j Q_{jk} n_k \biggr\}
\end{equation}
where $R_{jk} = \langle n_j n_k\rangle $ and $ \sum_{k} R_{jk} Q_{km} =
\delta_{jm}$.  This assumption is not critical to the approach,  but
makes for easier presentation and motivates the inner product used
below.   For stationary noise,  the kernel $Q_{jk}$ is determined by
the power-spectral density.

\subsection{Search Method}

Our approach is based on the excess-power search method
described in Refs.~\cite{Flanagan:1997sx,Anderson:2000yy} although extensions
of the present approach which might be more powerful will be described
elsewhere~\cite{Brady:2004}.  The gravitational wave signals form
a subspace $\cal{W}$ of the vector space ${\cal{V}}$ spanned by all
$H$-dimensional vectors given by Eq. (\ref{e:signal-vector}).   If
the subspace $\cal{W}$ is also a vector space,  then the optimal statistic for
the detection of these gravitational waves in Gaussian noise was derived by
Anderson \textit{et al.} \cite{Anderson:2000yy}.  One first decomposes
the signal into $\mathbf{s}_\parallel$ which lies in $\cal{W}$ and 
$\mathbf{s}_\perp$ which is orthogonal to $\cal{W}$,  so that
\begin{equation}
\mathbf{s} = \mathbf{s}_\perp + \mathbf{s}_\parallel \; .
\end{equation}
Then, the optimal statistic by which to identify gravitational-wave signals
which are in the vector space $\cal{W}$ is 
\begin{equation}
{\cal{E}} = (\mathbf{s}_\parallel, \mathbf{s}_\parallel)
\end{equation}
where the inner product is defined by
\begin{equation}
(\mathbf{a}, \mathbf{b}) = \sum_{i,j=0}^{N-1} a_i Q_{ij} b_j \; .
\label{e:inner-prod}
\end{equation}
The key to this approach is the determination of the signal
subspace $\mathcal{W}$ and the associated operator which projects
vectors from $\cal{V}$ into $\cal{W}$.   This was already pointed out by 
Anderson et al~\cite{Anderson:2000yy} where an example based on a
Fourier basis was described in detail.

\subsection{Constructing the signal subspace} 

Suppose numerical simulations yield a set of $M$ gravitational-wave
signals $\{h^I_j\}$ where $I=1,2,\ldots , M$.  Each waveform arises
from a simulation with some parameters ${\mathcal{P}}_I$.   If the
simulations are costly and only some representative sample can be
completed,  it is desirable to search for all signals with similar
characteristics to the sample.    A restricted search method might use the known waveforms
as matched filters to search for waveforms which match the simulations
well.  This approach is extremely limited when there are only a few
simulations or the simulations are known to be only qualitatively
correct.   The alternative approach, which we advocate here, is to use
the excess power method by constructing a vector space $\mathcal{W}$
of signals which capture the essential features of the waveforms given
by the simulations.   This approach also has its limitations since the
space of signals generated by some class of astrophysical sources is
not,  in general, expected to be a vector space.   Nevertheless,  the
projection of the detector output signal into a vector space of
sufficiently small dimensions can still be a powerful method to
distinguish signal from noise when combined with coincidence between
detectors.

The simplest approach to determine a vector subspace would use the
Gram-Schmidt method to determine an orthonormal set of vectors from
the waveforms $\{h^I_j\}$ and hence identify $\mathcal{W}$ as the
vector space spanned by all the simulation waveforms.   The
orthonormal basis of this space is therefore
\begin{equation}
e^I_j = \bar{h}^I_j / \sqrt{ (\bar{h}^I, \bar{h}^I) }
\end{equation}
where
\begin{equation}
 \bar{h}^I_j = h^I_j - \sum_{J=1}^{I-1} (h^I,e^J) e^J_j \; .
\end{equation}
Now,  any of the original simulation waveforms can be re-constructed
as a sum over the basis $h^I_j = \sum_{J} (h^I,e^J) e^J_j$.   

In general,  however,  we are interested not in the exact
re-construction of the original waveform,  but the identification of
robust features of the waveforms.   This is because the simulations may not be
completely accurate representations of the physical sources.   For
this reason,  we would like to use a smaller subspace which captures
the essential features of the simulated waveforms \emph{well enough}.
Our approach is similar to that used when determining the grid-spacing
in a matched filtering search and is most easily explained by
describing the algorithmic selection of basis vectors.

First,  normalize all the simulated waveforms using the inner product
(\ref{e:inner-prod});  denote the normalized waveforms by
$\hat{h}^I_j$.  Suppose then,  that one has used $K < M$ of the
simulated waveforms to construct an orthonormal set $\{ e^J_j \}$
where $J=1,\ldots,K$;  this is a basis for a vector space
${\mathcal{W}}_K \subset {\mathcal{W}}$.   Compute the norm
\begin{equation} 
\mu^L = || \sum_{J=1}^{K} (\hat{h}^L,e^J) e^J_j ||
\label{e:match}
\end{equation} 
for all waveforms which have not been used to construct
the basis of ${\mathcal{W}}_K$.
The quantity $\mu^L$ represents the \emph{match} of the $L$th
remaining waveform with the nearest vector in the space
${\mathcal{W}}_K$.   Select the signal with $\mu = min_L \mu^L$.
The number of basis vectors is incremented by one
if $\mu < \mu_{\mathrm{min}}$ where
$\mu_{\mathrm{min}}$ is some minimal match that will be tolerated in
the search.    This procedure is repeated  until a set of $M'
\leq M$ orthonormal vectors has been identified and $\mu^L >
\mu_{\mathrm{min}}$ for all remaining waveforms which have not been
used in the construction.   This approach will tend to select out a
subspace of signals $\mathcal{W}$ which capture the essential features
of the original set of simulated waveforms although the basis vectors
may be unphysical.

For typical source modeling simulations,  one expects that $M' < M$
since the waveforms produced for different parameter values often
share similar qualitative behavior.    As described above, 
the signal used to start the Gram-Schmidt construction is
chosen arbitrarily and might not result in the smallest vector space
of signals.   Thus,  the construction can be carried out repeatedly
selecting a different seed signal each time.  The seed signal which
leads to the smallest $M'$ is deemed the best choice for the
algorithm since the vector space of smallest dimension has been
identified.

It should be noted that waveforms sometimes need to be shifted in time
relative to each other in order to obtain the best match.   In each
step of the construction then,  it may be necessary to shift each of
the remaining waveforms in order to obtain the maximum match with the
subspace ${\mathcal{W}}_K$.   This is done in the examples
given below.

\section{Supernova searches as an example of the method}
\label{s:example}

The simulation of supernova explosions is extremely complicated
because it relies on detailed knowledge of physics at very high
densities in addition to strong field gravitational effects.    To
date,  only simple models of gravitational-wave production in core
collapse and supernova explosions have been
considered~\cite{Zwerger:1997,Dimmelmeier:2001rw,Ott:2003qg}.   These efforts provide
guidance to the gravitational wave astronomy community in the form of
waveform catalogs which represent a variety of observed behavior
depending on the parameters used in a particular simulation.

\subsection{Example of the Zwerger-M\"uller catalog}

The catalog of waveforms produced by Zwerger and M\"uller \cite{Zwerger:1997} have
become a reference point to calibrate searches for gravitational wave
bursts~\cite{Abbott:2003pb}.     The catalog provides the mass quadrupole wave
amplitudes $A^I(t)$ from axi-symmetric simulations of core-collapse.   The strain
at the detector is given in terms of these quadrupolar amplitudes by
\begin{equation} h^I(t) =
\frac{1}{8}\sqrt{\frac{15}{\pi}}sin^2{\theta}\frac{A^{I}(t)}{R}
\end{equation} 
where $R$ is the distance between the detector and the source and
$\theta$ is the {azimuthal angle}.  The catalog consists of 78
different waveforms arising from representative simulations across the
parameter space of the models.

\begin{figure}
{\includegraphics[width=0.9\textwidth]{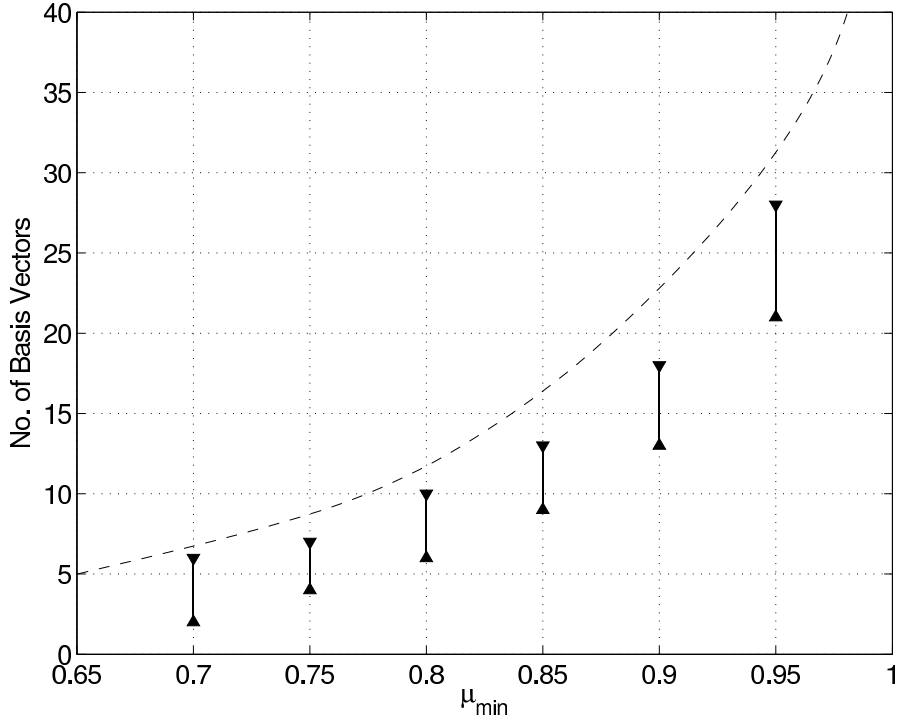}}
\caption{\label{f:basis-v-mismatch}
The number of basis vectors as a function of minimal match
$\mu_{\mathrm{min}}$ for
the ZM family of waveforms.   The dashed line shows the
results for the inner product appropriate to white noise.   The
triangles indicate the range of vlaues for the inner product 
appropriate to the LIGO-I design using different seed waveforms.
} 
\end{figure}

Following the method outlined in the previous section, we have
examined the number of basis vectors required to cover all 78
Zwerger-M\"uller (ZM) waveforms at various levels of minimal match
$\mu_{min}$.   The results are summarized in
Fig.~\ref{f:basis-v-mismatch} for a flat noise spectrum and the 
LIGO-I design spectrum~\cite{Barish:1999}.   At fixed
$\mu_{\mathrm{min}}$,  more basis vectors are generally needed for the
white noise case.   This is expected since high and low frequency
components of the waveforms are suppressed in the inner-product by the
LIGO-I noise spectrum.   There is some variation in the number of
basis vectors depending on the seed waveform used in the construction.
It is interesting to see that only 13 waveforms are needed to cover
the space of waveforms at $\mu_{\mathrm{min}}=0.9$.    This confirms
what can be seen by visually inspecting the ZM waveforms,
that is,  these waveforms have dominant features which are robust
across many examples in the catalog.  

\begin{figure}
\includegraphics[width=0.9\textwidth]{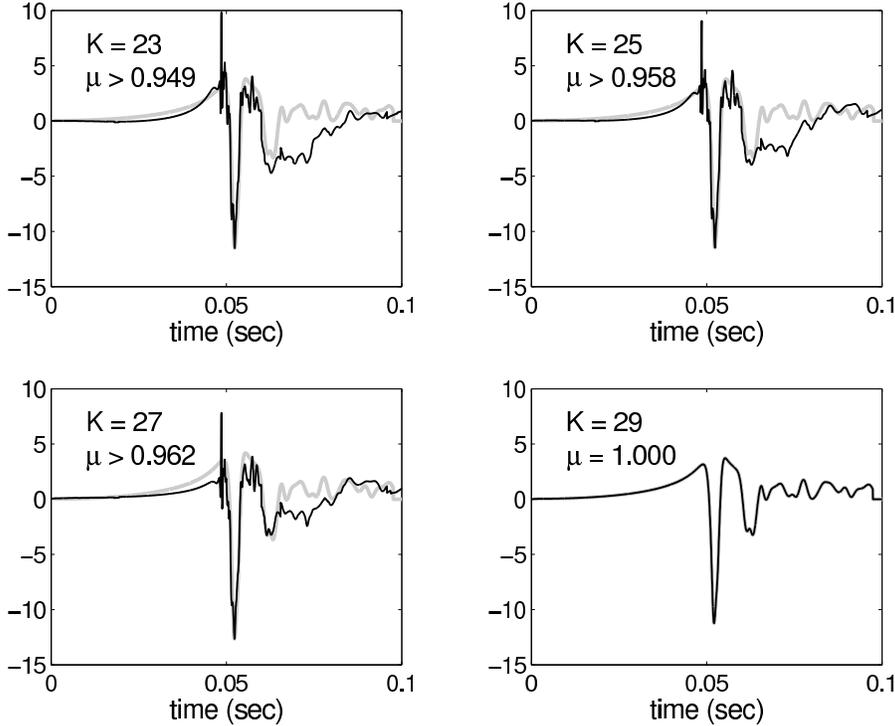}
\caption{\label{f:waveform-recon}
Waveform A3B3G2 from the ZM catalog (light-gray line)
shown along with reconstructed waveforms for a different numbers $K$
of basis vectors.   The value of the match between waveform A3B3G2 and
the corresponding vector space ${\cal{W}}_K$ is also indicated. The
fourth plot on the bottom right shows the exactly reconstructed
waveform once the  A3B3G2 is absorbed into the vector space ${\cal{W}}_K$.
}
\end{figure}

Figure~\ref{f:waveform-recon} shows waveform A3B3G2 from the
ZM catalog  plotted with the reconstructed waveform using
different numbers $K$ of basis vectors.   The match, as defined in
Eq.~(\ref{e:match}), between the
waveform and the vector space ${\cal{W}}_K$ is also shown.   As
indicated in Fig.~\ref{f:basis-v-mismatch},   the match is already
greater than $0.90$ for only $K=22$ basis vectors.    For
completeness,  the exactly reconstructed waveform is also shown 
when waveform A3B3G2 has been absorbed into the vector space
${\cal{W}}_K$.  

\subsection{Incorporating newer waveforms in the analysis}

A nice feature of this approach to using source simulation information
in searches for gravitational waves is the ease with which it can be
updated using new information.   Moreover,  the method provides an
algorithmic way to identify features of the simulated waveforms which
are robust from simulation to simulation.  

As a concrete example of this,  we consider the set of waveforms
produced by Ott et al. (OBLW)~\cite{Ott:2003qg}.  As with the ZM
simulations,  these waveforms also represent possible wave emission from core
collapse supernova explosions.   The simulations explore alternate
simulation methods in the numerical implementation and in
accounting for the complicated physics of supernovae.   When
seeking guidance for the detection of gravitational waves,  it is
interesting to understand how different the resulting waveform family
is from the ZM family.  This is important for two
reasons:  first, it tests the robustness of any search method that is
tuned to search for waveforms in the ZM family,  and
second,  it tests how much the waveform depends on changes in the
underlying physics of the source.

We determined the number of basis vectors needed to cover the 149
waveforms from both Zwerger-M\"uller and Ott et al. in the same way as
we determined the basis for only the ZM waveforms.   At
$\mu_{\mathrm{min}} = 0.9$,   one needs 33 basis vectors.   In this
example,  we used the seed waveform that minimized the number of
basis vectors for the ZM family alone.   The resulting basis was
constructed out of 12 waveforms from the ZM family and
21 from the OBLW family~\cite{Ott:2003qg};  only 6 of the 12 ZM
waveforms were used in the ZM-only basis set computed at this value of
$\mu_{\mathrm{min}}$.   For the purpose of this note,  the exact
numbers are not as important as the algorithmic approach
which has been taken.   

\section{Summary and future directions} 
\label{s:summary}

Searches for gravitational waves using broad-band detectors and the
extraction of physical insight from future detections will benefit
greatly from close interaction with scientists who model sources
of gravitational waves.   In the case where complete and accurate
information is available,  it is well known that matched filtering
provides the best approach to both detection and information
extraction.    For many physically complicated sources,  however,
there is a dearth of information available and even the best
simulations are not accurate enough to support matched filtering as
the main approach to detect the associated gravitational waves.   In this note,  we have
sketched an algorithmic approach to determining a vector space which
includes waveforms generated by the limited simulations and incorporates
waveforms of similar character.    The benefit
of this approach is that it is algorithmic and is easily updated using
newer information provided by simulations.

The supernova simulations of Zwerger-M\"uller~\cite{Zwerger:1997} and
Ott et al.~\cite{Ott:2003qg} have been used as an example to demonstrate
the method.  In a future publication,   we will present more detailed
investigations of these and other supernova simulations along with
estimates of the performance of the approach in the presence of noise.
This is particularly important since the spiky nature of the basis
vectors could increase the false alarm probability beyond a tolerable
level in real detector noise

One of the more interesting avenues to explore with this approach is,
however, the detection of the late stages of binary black hole
inspiral and merger~\cite{BRY}.   Current efforts in this direction
have focused on identifying a detection family of waveforms for use in
the search~\cite{BCV:2003a};   given the large number of templates required
to span known waveform models,  it seems worthwhile to explore the
method outlined here as an alternative.   

\ack

We would like to thank Duncan Brown,  Jolien Creighton and Alan Wiseman for useful
discussions.  This work has been supported by the National Science
Foundation Grant PHY-0200852.  Patrick Brady is also grateful to the
Alfred P Sloan Foundation and the Research Corporation Cottrell
Scholars Program for support.


\section*{References}
\begin{thebibliography}{99}

\bibitem{wainstein:1962}
L.~A. Wainstein and V.~D. Zubakov, \textit{Extraction of signals from noise}
  (Prentice-Hall, London, 1962)

\bibitem{Blanchet:1996pi}
L.~Blanchet, B.~R.~Iyer, C.~M.~Will and A.~G.~Wiseman
1996
\emph{Class.\ Quant.\ Grav.}  {\bf 13} 575
[arXiv:gr-qc/9602024]

\bibitem{Allen:1999yt}
B. Allen {\it et~al.} 1999 \emph{Phys. Rev. Lett.} {\bf 83},  1498
[arXiv:gr-qc/9903108]

\bibitem{Tagoshi:2000bz}
H.~Tagoshi {\it et al.}  (TAMA Collaboration)
2001
\emph{Phys.\ Rev.\ D} {\bf 63} 062001
[arXiv:gr-qc/0012010]

\bibitem{Abbott:2003pj}
B.~Abbott {\it et al.}  (The LIGO Scientific Collaboration)
Analysis of LIGO data for gravitational waves from binary neutron stars
\emph{Preprint}
[arXiv:gr-qc/0308069].

\bibitem{Cutler:ys}
C.~Cutler and E.~E.~Flanagan,
Phys.\ Rev.\ D {\bf 49} (1994) 2658
[arXiv:gr-qc/9402014].

\bibitem{Finn:1992xs}
L.~S.~Finn and D.~F.~Chernoff,
Phys.\ Rev.\ D {\bf 47} (1993) 2198
[arXiv:gr-qc/9301003].


\bibitem{Cutler:2002me}
C.~Cutler and K.~S.~Thorne,
in \emph{Proceedings of 
General Relativity and Gravitation XVI}, edited by Nigel T. Bishop 
and Sunil D. Maharaj (Singapore, World Scientific, 2002).
arXiv:gr-qc/0204090.

\bibitem{Flanagan:1997sx}
E.~E.~Flanagan and S.~A.~Hughes,
Phys.\ Rev.\ D {\bf 57} (1998) 4535
[arXiv:gr-qc/9701039].

\bibitem{Zwerger:1997}
T.~Zwerger and E.~M\"uller
Astron. Astrophys. p. 209 (1997).

\bibitem{Dimmelmeier:2001rw}
H.~Dimmelmeier, J.~A.~Font and E.~M\"uller,
Astrophys.\ J.\  {\bf 560} (2001) L163
[arXiv:astro-ph/0103088].

\bibitem{Ott:2003qg}
C.~D.~Ott, A.~Burrows, E.~Livne and R.~Walder,
Astrophys.\ J.\  {\bf 600} (2004) 834
[arXiv:astro-ph/0307472].

\bibitem{Lehner:2001wq}
L.~Lehner,
Class.\ Quant.\ Grav.\  {\bf 18} (2001) R25
[arXiv:gr-qc/0106072].

\bibitem{Baker:2002qf}
J.~G.~Baker, M.~Campanelli, C.~O.~Lousto and R.~Takahashi,
Phys.\ Rev.\ D {\bf 65} (2002) 124012
[arXiv:astro-ph/0202469].

\bibitem{Bruegmann:2003aw}
B.~Bruegmann, W.~Tichy and N.~Jansen,
arXiv:gr-qc/0312112.


\bibitem{Nicholson:1996ys}
D. Nicholson {\it et~al.}, Phys. Lett. {\bf A218},  175  (1996).

\bibitem{IGEC:2000}
{Z.~A. Allen} {et~al.}, ({International Gravitational Event Collaboration}),
{Phys. Rev. Lett.} \textbf{{85}}, {5046} ({2000}).


\bibitem{Abbott:2003pb}
B.~Abbott {\it et al.}  [LIGO Scientific Collaboration],
``First upper limits from LIGO on gravitational wave bursts,''
to appear in Phys.\ Rev.\ D 
[arXiv:gr-qc/0312056].

\bibitem{igec:2003}
{P.}~{Astone} {et~al.},
  ({International Gravitational Event Collaboration}),
  {Phys. Rev. D} \textbf{{68}},
  {022001} ({2003}), {[arXiv:astro-ph/0302482]}.

\bibitem{Anderson:2000yy}
W.~G.~Anderson, P.~R.~Brady, J.~D.~E.~Creighton and E.~E.~Flanagan,
Phys.\ Rev.\ D {\bf 63} (2001) 042003
[arXiv:gr-qc/0008066].

\bibitem{Brady:2004}
P.~R.~Brady and S.~Ray-Majumdar,  ``Interfacing gravitational-wave 
source modelling and observations'',
in preparation (2004).

\bibitem{Barish:1999}
{{B.~C.} {Barish}} {and} {{R.}~{Weiss}},
  {Phys.\ Today} \textbf{{52 (Oct)}},
  {44} ({1999}).

\bibitem{BRY}
P.~R.~Brady,  S.~Ray-Majumdar and P. Yu, 
in preparation (2004).

\bibitem {BCV:2003a}
{{A.}~{Buonanno}}, {{Y.}~{Chen}}, {and} {{M.}~{Vallisneri}},
{Phys. Rev. D} \textbf{{67}}, {024016} ({2003}).

\end{thebibliography}
\end{document}